\documentclass{article}
\usepackage{amsmath}
\usepackage{makeidx}
\usepackage{amssymb}

\newtheorem{theorem}{Theorem}

\newtheorem{remark}[theorem]{Remark}

\begin{document}

\title{Surprising Structures Hiding at Penrose's Future Null Infinity \ \ \ }
\author{Ezra .T. Newman\footnote{University of Pittsburgh}}
\date{1.25.17}
\maketitle
\begin{abstract}
Since the late1950s, almost all discussions of Asymptotically Flat
(Einstein-Maxwell) Space-Times have taken place in the context of Penrose's
Null Infinity, $\mathcal{I}^{+}.$\ $\ $In addition,\ almost all calculations
have used the Bondi coordinate and tetrad systems. \ We show - first, that
there are other natural coordinate systems, near $\mathcal{I}^{+},$
(analogous to light-cones in flat-space) that are based on (asymptotically)
shear-free null geodesic congruences (analogous to the flat-space case). \
Using these new coordinates and their associated tetrad, we \textit{define
the complex dipole moment, i.e., as the mass dipole plus i times angular
momentum,} from the $l=1,\ $harmonic coefficient of a component of the
asymptotic$\ $Weyl tensor. Second, from this definition, from the Bianchi
Identities and from the Bondi mass and linear momentum, we show that \ there
exists a large number of results - identifications and dynamics - identical
to those of classical mechanics and electrodynamics. They include, among
many others, \textbf{P}=M\textbf{v}+..., \textbf{L}=\textbf{r}x\textbf{P}, \
spin, Newtons 2nd Law with the Rocket force term (\.{M}\textbf{v}) and
radiation reaction, angular momentum conservation and others. \ All these
relations take place in the rather mysterious H-Space.

This leads to the enigma: "why do these well known relations of classical
mechanic take place in $H-$space?" and "What is the physical meaning of $H-$%
space?"
\end{abstract}

\section{ Introduction}

The modern era of the study of Gravitational Radiation began in the 1950s
with the pioneering work of Hermann Bondi\cite{Bondi}. This was quickly
expanded by the major contributions of Rainer Sachs\cite{Sachs} and Roger
Penrose\cite{Penrose 1}\cite{NP}\cite{NU}\cite{Scholarpedia} among many
others. \ After years of further developments, theoretical, numerical and
observational, we had its culmination with the observation and analytic
understanding of the collision and merger of the pair of black holes that
produced the gravitational wave signal, GW105,\cite{GW}, that was seen by
LIGO in 2016. Gravitational wave theory now could play a major role in
astrophysics and physics.

Bondi's work began with integrating the Einstein Equations in the asymptotic
region - in the vicinity of future null infinity. \ This involved the
important step of using special null surfaces as part of the coordinate
system referred to as Bondi coordinates, ($r,u$,$\zeta ,\overline{\zeta }$).
(The Bondi coordinates are defined uniquely up to a group of transformations
known as the BMS group.\cite{Penrose 1}\cite{ANK}) The idea of working near
or even at \textit{infinity}, (though at the beginning was slightly
nebulous), was formalized (by Penrose\cite{Penrose 1}) by bringing infinity
into a finite region of the space-time by its\ conformal compactification,
(rescaling and contraction). Future null Infinity was then represented by a
null three-surface in space-time (referred to as$\ \mathcal{I}^{+},\ $%
vocalized by SCRI+). \ $\mathcal{I}^{+}$,$\mathcal{\ }$a null 3-surface with
the topology of $\mathcal{S}^{2}\times \mathcal{R}$,$\mathcal{\ }$%
(visualized as a light-cone at future null infinity, apex at time-like
infinity) is$\mathcal{\ }$coordinatized by the complex stereographic
coordinates, ($\zeta =e^{i\phi }\cot \frac{\theta }{2},\overline{\zeta }%
=e^{-i\phi }\cot \frac{\theta }{2}$)$,\ $on the $\mathcal{S}^{2}\ \ $part$,\ 
$(labeling the null generators) and with $u$ on \nolinebreak the$\ \mathcal{%
R\ }$part (labeling the cross-sections of $\mathcal{I}^{+}\mathcal{)}$. \
The BMS group can be described as coordinate transformations among the
coordinates\ ($u$,$\zeta ,\overline{\zeta }).$ In addition to the
introduction of null surfaces, Bondi's other insight was the realization
that several components of the asymptotic Weyl tensor could be identified
with\ the \textit{\ mass/energy} and \textit{linear momentum of the} \textit{%
source }and \textit{their loss} - in analogy, in Maxwell theory, to \textit{%
charge and charge conservation} as integrals of the fields at infinity. \
These Weyl tensor components come from the harmonic components of the
leading coefficients in the $r^{-1}\ ~$expansion in the spin-coefficient
version of the Weyl tensor and are thus functions just on $\mathcal{I}^{+},\ 
$i.e., are functions of ($u$,$\zeta ,\overline{\zeta }$).

\bigskip An important idea to recognize is that the leading far-field
components of the Weyl tensor (in the spin-coefficient formulation\cite%
{Scholarpedia}), depend very much on the choice of the tetrad and
coordinates systems to be used on $\mathcal{I}^{+}.\ \ $In the past almost
all Weyl tensor components were chosen in a Bondi system. However, in the
present work, the major new ingredient that leads to our results is the
introduction of totally new coordinates and tetrad systems on $\mathcal{I}%
^{+}\ $and its neighborhood,\ (very different from Bondi's)\ that very
closely mimic certain natural coordinate systems on the Minkowski space $%
\mathcal{I}^{+}.\ \ $This allows us to describe other functions on $\mathcal{%
I}^{+}\ $(arising from other Weyl$\ $tensor coefficients)\ that yield - with
the Bianchi identities - a variety of additional physical quantities, such
as angular momentum, center of mass, its position and velocity, their
evolution, as well as force laws and electric and magnetic dipole moments
and center of charge. In other words these results and relationships from
classical mechanics simply appear as components of the Weyl tensor at
infinity.

The prime idea involved in the choice of these systems is use of the \textit{%
special null surfaces} that are associated with asymptotically \textit{%
shear-free null geodesic congruences \ - }as directly opposed to Bondi's
null surfaces which do have \textit{non-vanishing} asymptotic shear - the 
\textit{time}-\textit{integral }of the Bondi news function.

\qquad These special null surfaces, which are standard and easily understood
surfaces in Minkowski space, are of two types. The first are the null cones
(with generators automatically shear-free and twist-free) with apex on
arbitrary time-like curves - a special case being time-like geodesics. The
second type arise (formally) from complex light-cones with apex on complex
world-lines\cite{ANK}\cite{AN}. In this case, the associated real
congruence, though shear-free, is now twisting, They will be first reviewed
and described in Sec. II. Sec.III will be devoted to their generalization
(in asymptotically flat spaces) to\textit{\ asymptotically shear-free
congruences}, i.e., their definition and construction. \ Though at first it
appears that there is a serious impediment to their construction, it turns
out that by a slight zigzag or maneuver, the impediment can be overcome and
the construction can be completed in exact analogy to the flat-space case. \
We will have coordinate systems on (asymptotically flat) $\mathcal{I}^{+}$,
very closely matching those in Minkowski space. The \textit{asymptotic
generators (the null geodesics)} of the complex surfaces (by construction)
will be \textit{real }and asymptotically \textit{shear-free} but, in
general, they will be twisting. In Secs.IV and V we will, by using earlier
work, show how the asymptotic Weyl tensor components, expressed in the new
coordinates with their associated tetrads, naturally yield\ a large number
of functions determining the interior space time properties as mentioned
earlier. \ In the discussion, the strange appearance of $H$%
-space-coordinates is addressed. \ We emphasize that though complex ideas
are used, we are dealing with real space-times.

For close to 50 years the coordinatization of $\mathcal{I}^{+}\ $by Bondi
coordinates has been almost sacrosanct \ - nevertheless we present (what we
believe is a strong argument) that other choices of coordinates on $\mathcal{%
I}^{+}\ $have considerable value and their use should be seriously
considered.

\section{$\mathcal{I}^{+}\ $of Minkowski space}

Using standard Minkowski coordinates, $x^{a},\ $with metric $\eta _{ab}\ $%
and signature (+,-,-,-), the family of null cones with apex at the origin is
described parametrically, ($u,r$,$\zeta ,\overline{\zeta }$), by

\begin{equation}
x^{a}=ut^{a}+r\widehat{l}^{a}(\zeta ,\overline{\zeta })
\label{coordinate trans.I}
\end{equation}%
with $\widehat{l}^{a}(\zeta ,\overline{\zeta })\ $a null vector that sweeps
out the null cone\ as $\zeta =\cot \frac{\theta }{2}e^{i\phi }\ $varies over
the sphere of null directions, i.e.,%
\begin{eqnarray}
\widehat{l}^{a}(\zeta ,\overline{\zeta }) &=&\frac{\sqrt{2}}{2}{\Large (}1,%
\frac{\zeta +\overline{\zeta }}{1+\zeta \overline{\zeta }},-i\frac{\zeta -%
\overline{\zeta }}{1+\zeta \overline{\zeta }},-\frac{1-\zeta \overline{\zeta 
}}{1+\zeta \overline{\zeta }}{\large )}=(\frac{\sqrt{2}}{2},\frac{1}{2}%
Y_{1i}^{0})  \label{description} \\
t^{a} &=&\delta _{0}^{a}.  \label{unit}
\end{eqnarray}

The $r\ \ $is\ the affine parameter along the null generators of the cone
and $u\ $the time at the spatial origin (or the retarded time on the cone
itself).

An alternate interpretation of Eq.(\ref{coordinate trans.I}) is that it is
the coordinate transformation between the $x^{a}\ $and the null coordinates (%
$u,r$,$\zeta ,\overline{\zeta }$). \ We will refer to the coordinate
transformation (and coordinates) given by Eq.(\ref{coordinate trans.I}), as
well as its later generalization in Sec.III, as "static null coordinates".

\textbf{Aside}: \ The full null tetrad, ($\widehat{l},\widehat{n},\widehat{m}%
,\overline{\widehat{m}}$), associated with Eq.(\ref{description}), is given
by%
\begin{eqnarray}
\widehat{n}^{a} &=&\frac{\sqrt{2}}{2P}(1+\zeta \overline{\zeta },-(\zeta +%
\overline{\zeta }),\text{ }i(\zeta -\overline{\zeta }),1-\zeta \overline{%
\zeta }),  \label{three} \\
\widehat{m}^{a} &=&\eth l^{a}=\frac{\sqrt{2}}{2P}(0,1-\overline{\zeta }%
^{2},-i(1+\overline{\zeta }^{2}),\text{ }2\overline{\zeta }),  \notag \\
\overline{\widehat{m}}^{a} &=&\overline{\eth }l^{a}=\frac{\sqrt{2}}{2P}%
(0,1-\zeta ^{2},\text{ }i(1+\zeta ^{2}),2\zeta ),  \notag \\
P &=&1+\zeta \overline{\zeta }.  \label{P}
\end{eqnarray}

The metric tensor in these new coordinates, given by%
\begin{equation}
ds^{2}=du^{2}+2dudr-4r^{2}P^{-2}d\zeta d\overline{\zeta }
\end{equation}%
can be conformally transformed (rescaled) by $\Omega ^{2}=r^{-2},\ $leading
to%
\begin{equation}
d\widetilde{s}^{2}=\widetilde{g}_{ab}dx^{a}dx^{b}=\Omega ^{2}du^{2}+\Omega
^{2}2dudr-4P^{-2}d\zeta d\overline{\zeta }.
\end{equation}

The surface defined by $\Omega =0,$ a null surface$,\ $is\ identified as the 
$\mathcal{I}^{+}\ $of Minkowski space$.\ $It can be thought of as the
intersection of the endpoints of the future null cones that have apex on the
world-line $x^{a}=ut^{a},\ $with$\ $future\ null infinity. It is
coordinatized by$\ (u,\zeta ,\overline{\zeta })\ $and is a special case of
Bondi coordinates.

These null coordinates can be generalized to a new set, ($t,r^{\ast },\zeta
^{\ast },\overline{\zeta }^{\ast }$), by basing them on null cones with apex
on an arbitrary time-like world-line, $x^{a}=\xi ^{a}(t),\ $by\ the $\ $%
parametric\ form, ($t,\zeta ^{\ast },\overline{\zeta }^{\ast }$),\ $t\ $a
real parameter, 
\begin{equation}
x^{a}=\xi ^{a}(t)\ +r^{\ast }\widehat{l}^{\ast a}(\zeta ^{\ast },\overline{%
\zeta }^{\ast }).  \label{coordinate transf.II}
\end{equation}

\noindent\ $\widehat{l}^{\ast a}\ \ $is again a null vector sweeping out the
null directions on the cone. By equating the right sides of Eqs.(\ref%
{coordinate trans.I})\ and$\ $(\ref{coordinate transf.II})$,\ $multiplying
by the four null tetrad vectors associated with $\widehat{l}^{a}(\zeta ,%
\overline{\zeta }),\ $Eq.(\ref{three}),$\ $and passing to the limit $%
r=>\infty ,\ ($or $\Omega =0)\ $we find the relationship, (the light-cone
cuts)\cite{Kent}\cite{KN}, 
\begin{equation}
u=G_{F}(t,\zeta ,\overline{\zeta })\equiv \xi ^{a}(t)\ l_{a}(\zeta ,%
\overline{\zeta })=\frac{\sqrt{2}}{2}\xi ^{0}(t)+\frac{1}{2}\xi
^{i}(t)Y_{1i}^{0}(\zeta ,\overline{\zeta }),  \label{flat cuts}
\end{equation}

\noindent that describes, in Bondi coordinates, the intersection of the null
cones, apex on $\xi ^{a}(t),\ $with $\mathcal{I}^{+}$. \ Eq.(\ref{flat cuts}%
), thus \underline{\textit{defines a one real parameter}}, $t,\ $family of
'slicings' of $I^{+}.$

At a point on $\mathcal{I}^{+},\ (u,\zeta ,\overline{\zeta }),\ $where the
two null vectors $\widehat{l}^{\ast a}\ $and $\widehat{l}^{a}\ $meet, the
null angle (on their past light-cone) between them (stated in stereographic
coordinates, $L\ $and $\overline{L}\ ),$\ is given by 
\begin{equation}
L=\eth G_{F},\ \ \overline{L}=\overline{\eth }G_{F}\   \label{L}
\end{equation}

\noindent Asymptotically, the two null vectors (and associated tetrads), $%
\widehat{l}^{\ast a}\ $and $\widehat{l}^{a},$ are related by $\ $%
\begin{eqnarray}
\widehat{l}^{\ast a} &=&\widehat{l}^{a}+b\widehat{\overline{m}}^{a}+%
\overline{b}\widehat{m}^{a}+b\overline{b}\widehat{n}^{a},
\label{null rotation} \\
\widehat{m}^{\ast a} &=&\widehat{m}^{a}+b\widehat{n}^{a},  \notag \\
\widehat{n}^{\ast a} &=&\widehat{n}^{a},  \notag \\
b &=&-\frac{L}{r}+0(r^{-2}).  \notag
\end{eqnarray}

It is useful to distinguish between the null coordinates based Eq.(\ref%
{coordinate trans.I}), ($u,r$,$\zeta ,\overline{\zeta }$), referred to as
static null coordinates and those based on Eq.(\ref{coordinate transf.II}), (%
$t,r^{\ast },\zeta ^{\ast },\overline{\zeta }^{\ast }$),$\ $referred to as
dynamic or comoving null coordinates. \ 

An important point for us is to note that the cuts $u=G_{F}(t,\zeta ,%
\overline{\zeta })\ $satisfy the (so-called) \textit{flat-space good cut
equation},%
\begin{equation}
\eth ^{2}G_{F}=0,
\end{equation}%
namely the condition for the null normals to the 'cuts' to define null
vectors that are \textit{shear-free\cite{ANK}\cite{Penrose 1}\cite%
{Scholarpedia}. }In the following section dealing with asymptotically flat
spaces, this equation will be generalized to the \textit{good cut equation,}%
\begin{equation}
\eth ^{2}G=\sigma ^{0}(G,\zeta ,\overline{\zeta })  \label{GCEq}
\end{equation}%
where $\sigma ^{0}(u,\zeta ,\overline{\zeta })\ $is the asymptotic shear
(the time integral of the Bondi news function).

\begin{remark}
\bigskip For relevance and analogy with the following section, we point out
that in this flat space discussion we \underline{could have taken} the $\xi
^{a}(\tau )\ $to be a complex world-line\cite{ANK}\cite{AN}. The $L=\eth
G_{F}\ $and its complex conjugate, via Eq.(\ref{null rotation}), would still
lead to $\widehat{l}^{\ast a}$ being shear-free but now it would be
twisting.\ \ The cuts however would be intrinsically complex and their real
parts would have to be chosen - but only after the differentiation.
\end{remark}

\section{Sec.III, $\mathcal{I}^{+}\ $of Asymptotically Flat Space}

Turning from Minkowski space to asymptotically flat spaces, we begin with $%
\mathcal{I}^{+}\ $constructed from fixed but arbitrary Bondi coordinates, ($%
u,r,\zeta ,\overline{\zeta }$), and Bondi tetrad, $(l^{a},m^{a},\overline{m}%
^{a},n^{a}),\ $with$\ l^{a}$\ tangent to the Bondi null surfaces, $m^{a},%
\overline{m}^{a},\ $\ tangent to the Bondi cuts at\ $\mathcal{I}^{+}\ $and $%
n^{a}\ \ $tangent to the $\mathcal{I}^{+}\ $null generators, with $n,m,%
\overline{m}\ $\ parallel propagated down the null geodesics on $u$. \ The
radiation free-data is 
\begin{equation*}
\sigma ^{0}(u,\zeta ,\overline{\zeta })=\xi ^{ij}(u)Y_{2ij}^{2}+....
\end{equation*}%
$\ $

We now mimic the construction of the shear-free congruences of the previous
section.

\ It is known\cite{ANK}\cite{New} that the general regular solution of Eq.(%
\ref{GCEq}) depends on four complex parameters,\ $z^{a},\ $(defining $H$%
-space) that can be taken as functions of the \textit{complex parameter} $%
\tau \ \ $and written as $z^{a}=\xi ^{a}(\tau ),.\ $i.e.,\ as an arbitrary
(to be determined) complex world-line in $H$-space$.\ \ $The solution (via
coordinate conditions on the first four harmonics,$\ l=0,1$) can be written
in the form%
\begin{eqnarray}
u &=&G^{\ast }(\tau ,\zeta ,\overline{\zeta })\equiv G(\xi ^{a}(\tau ),\zeta
,\overline{\zeta })\equiv z^{a}l_{a}(\zeta ,\overline{\zeta })+\xi
^{ij}(z^{a})Y_{ij}^{2}(\zeta ,\overline{\zeta })+...  \label{cuts} \\
z^{a} &=&\xi ^{a}(\tau )  \label{world-line}
\end{eqnarray}%
with the quadrupole term $\xi ^{ij},\ $arising from the data, $\sigma
^{0}=\xi ^{ij}Y_{ij}^{2}(\zeta ,\overline{\zeta })+..$. From the freedom to
rescale $\tau ,\ $i.e.,$\ \tau ^{\ast }=F(\tau ),\ $we set $\xi ^{0}(\tau
)=\tau .\ $\ This is the velocity normalization and the slow motion
approximation, with$\ \xi ^{a\ \prime }\equiv v^{a},\ \xi ^{0\ \prime }=%
\sqrt{1+v^{i2}}\approx 1\ .\ \ $Eq.(\ref{cuts}) then becomes%
\begin{equation}
u=\frac{\tau }{\sqrt{2}}-\frac{1}{2}\xi ^{i}(\tau )Y_{1i}^{0}(\zeta ,%
\overline{\zeta })+\xi ^{ij}(\xi ^{a}(\tau ))Y_{ij}^{0}(\zeta ,\overline{%
\zeta })+..  \label{cuts*}
\end{equation}

(The $\xi ^{ij}\ $turn out to be the time-derivatives of the gravitational
quadrupoles:

\begin{equation*}
\xi ^{ij}=\frac{\sqrt{2}G}{24c^{4}}(Q_{Mass}^{ij\prime \prime
}+iQ_{spin}^{ij\prime \prime }).
\end{equation*}

The idea is now to generalize the flat-space cuts, Eq.(\ref{flat cuts}), 
\textit{to a one-parameter family of real cuts} in the asymptotically flat
situation, via Eq.(\ref{cuts*}). Unfortunately this does not work
immediately since,\textit{\ in general}, for arbitrary $\sigma ^{0}(u,\zeta ,%
\overline{\zeta }),\ $the $G^{\ast }(\tau ,\zeta ,\overline{\zeta })\ \ $%
will be complex and there will essentially be no real cuts. \ Before we see
a way around this problem we mention that IF the $\sigma ^{0}\ \ $was of
pure electric type\cite{NP2} then real cuts could be found and the situation
would resemble Eq.(\ref{flat cuts}). \ For general type of $\sigma ^{0}\ $%
the \textbf{Remark }of the previous section becomes relevant, i.e., it is
the analogue of the present case.

\ The way around the problem of the complexity of the cuts, Eq. (\ref{cuts},
is the following: treating $\tau \ $as complex, we first construct the null
angles, 
\begin{eqnarray}
L &=&\eth G(\xi ^{a},\zeta ,\overline{\zeta }),  \label{L&Lbar} \\
\ \overline{L} &=&\overline{\eth }\overline{G}(\overline{\xi }^{a},\zeta ,%
\overline{\zeta }),\ \   \notag
\end{eqnarray}%
to produce an \textit{asymptotically shear-free null vector field}, $l^{\ast
a},\ $via Eq.(\ref{null rotation}),

\begin{eqnarray}
l^{\ast a} &=&l^{a}+b\overline{m}^{a}+\overline{b}m^{a}+0(r^{-2}),
\label{shear-free} \\
b &=&-\frac{L}{r}+0(r^{-2}).  \notag
\end{eqnarray}

Note: We use$\ \overline{L}\ \ $and do not use $\widetilde{L}=\overline{\eth 
}G(\xi ^{a},\zeta ,\overline{\zeta }).$

\ Next, using 
\begin{eqnarray}
\tau &=&t+i\lambda ,  \label{t,lambda} \\
\xi ^{a}(\tau ) &=&\xi _{R}^{a}(t,\lambda )+i\xi _{I}^{a}(t,\lambda ),
\label{DECOMPOSE}
\end{eqnarray}%
we decompose $G^{\ast }(\tau ,\zeta ,\overline{\zeta })\ $into its real and
imaginary parts,%
\begin{equation*}
G(\xi ^{a}(\tau ),\zeta ,\overline{\zeta })=G_{R}(\xi _{R}^{a}(t,\lambda
),\xi _{I}^{a}(t,\lambda ),\zeta ,\overline{\zeta })+iG_{I}(\xi
_{R}^{a}(t,\lambda ),\xi _{I}^{a}(t,\lambda ),\zeta ,\overline{\zeta }).
\end{equation*}

By setting $G_{I}=G_{I}(\xi _{R}^{a}(t,\lambda ),\xi _{I}^{a}(t,\lambda
),\zeta ,\overline{\zeta })=0\ $and$\ $solving it for\ $\lambda ,$\ i.e., 
\begin{equation}
\lambda =\Lambda (t,\zeta ,\overline{\zeta }),  \label{LAMBDA}
\end{equation}%
and substituting $\lambda \ $into $G_{R}\ ,\ $we have the\textit{\ 
\underline{\textit{one real parameter family of real cuts}},}%
\begin{equation}
u=G_{R}(\xi _{R}^{a}(t,\Lambda ),\xi _{I}^{a}(t,\Lambda ),\zeta ,\overline{%
\zeta }).  \label{real cuts}
\end{equation}

This construction can be done under fairly general conditions assuming $%
\partial G_{I}/\partial \lambda \neq 0.$\qquad \qquad

With the $L\ $and $\overline{L},\ $of\ Eqs.(\ref{L&Lbar}) and (\ref%
{shear-free})\ \textit{evaluated on the real cuts}, the associated \textit{%
real but twisting shear-free null vector field\ }is,%
\begin{equation}
l^{\ast a}=l^{a}-\frac{L}{r}\overline{m}^{a}-\frac{\overline{L}}{r}%
m^{a}+0(r^{-2}).  \label{l*}
\end{equation}

We now have the situation in asymptotically flat spaces that is \textit{%
totally analogous to the situation we had in Minkowski space}. We refer to
the $H$-space coordinates,$\ z^{a},\ $as complex \textit{pseudo-Minkowski}
coordinates, since in the flat-limit they are the complexified Minkowski
coordinates. \ The $\mathcal{I}^{+}\ $coordinates associated with these 
\textit{pseudo-Minkowski} coordinates i.e., the ($\tau ,\zeta ,\overline{%
\zeta }$), will be referred to as pseudo-Minkowski cuts.

For the reality considerations, we have two separate cases: (1) \ the
comoving choice of an arbitrary $\xi ^{a}(\tau ),\ $in Eq.(\ref{cuts*}),$\ $%
to be determined by choosing it to be the complex center of mass or (2) by
the choice of the static pseudo-Lorentzian $\mathcal{I}^{\mathcal{+}}\ $%
coordinates, i.e., by taking (a `straight' $H$-space world-line) $\xi ^{a}=%
\widehat{\tau }\delta _{0}^{a},\ $for the cuts:%
\begin{equation}
u=\frac{\widehat{\tau }}{\sqrt{2}}+\xi ^{ij}(\xi ^{a}(\widehat{\tau }%
))Y_{ij}^{2}(\zeta ,\overline{\zeta })+..  \label{cuts**}
\end{equation}%
$\ \ $

If we assume that both $\lambda \ $is small and the slow motion
approximation, these constructions can be \textit{explicitly} carried out. \
They lead, via the assumed form of the \textit{asymptotic shear}, 
\begin{equation}
\sigma ^{0}=\xi ^{ij}(u)Y_{2ij}^{2}(\zeta ,\overline{\zeta })+...,
\label{asym shear}
\end{equation}%
to the linearized expressions (that we need), 
\begin{eqnarray}
&&  \notag \\
&&\text{case }1  \label{case`1} \\
\lambda  &=&\Lambda (t,\zeta ,\overline{\zeta })\equiv \frac{\sqrt{2}}{2}\xi
_{I}^{i}(t)Y_{1i}^{0}(\zeta ,\overline{\zeta })-\sqrt{2}\xi
_{I}^{ij}(t)Y_{2ij}^{0}(\zeta ,\overline{\zeta }),  \label{lambda} \\
u_{R} &=&G_{R}=\frac{t}{\sqrt{2}}-\frac{1}{2}\xi _{R}^{i}(t)Y_{1i}^{0}(\zeta
,\overline{\zeta })+\xi _{R}^{ij}(t)Y_{2ij}^{0}(\zeta ,\overline{\zeta }),
\label{u_R} \\
\xi ^{ij}(u) &=&\xi _{R}^{ij}(u)+i\xi _{I}^{ij}(u).  \label{Quad}
\end{eqnarray}

\begin{eqnarray}
&&\text{case }2  \label{case 2} \\
\widehat{\tau } &=&\widehat{t}+i\widehat{\lambda } \\
\widehat{\lambda } &=&\widehat{\Lambda }(\widehat{t},\zeta ,\overline{\zeta }%
)\equiv -\sqrt{2}\xi _{I}^{ij}(\widehat{t})Y_{2ij}^{0}(\zeta ,\overline{%
\zeta }),  \label{lambda*} \\
u_{R} &=&G_{R}=\frac{\widehat{t}}{\sqrt{2}}+\xi _{R}^{ij}(\widehat{t}%
)Y_{2ij}^{0}(\zeta ,\overline{\zeta }).  \label{u*}
\end{eqnarray}

\section{Review and Further Developments}

As mentioned earlier, much of the material described here will involve
functions or structures that `live' on $\mathcal{I}^{+}$ \ and originate
with the leading Weyl and Maxwell tensor components. Our major interest will
center on the asymptotic behavior, the physical meaning, the evolution and
transformation properties of these tensors. \ Using Bondi coordinates and
tetrad, the five complex self-dual NP components of the Weyl tensor and
three complex Maxwell components are\cite{NP}:

\begin{eqnarray}
\Psi _{0} &=&-C_{abcd}l^{a}m^{b}l^{c}m^{d}=-C_{1313},  \label{W0} \\
\Psi _{1} &=&-C_{abcd}l^{a}n^{b}l^{c}m^{d}=-C_{1213},  \label{W1} \\
\Psi _{2} &=&-C_{abcd}l^{a}m^{b}\overline{m}^{c}n^{d}=-C_{1342},  \label{W2}
\\
\Psi _{3} &=&-C_{abcd}l^{a}n^{b}\overline{m}^{c}n^{d}=-C_{1242},  \label{W3}
\\
\Psi _{4} &=&-C_{abcd}n^{a}\overline{m}^{b}\overline{m}^{c}n^{d}=-C_{2442}.
\label{W4}
\end{eqnarray}

\begin{eqnarray*}
\phi _{0} &=&F_{ab}l^{a}m^{b}, \\
\phi _{1} &=&\frac{1}{2}F_{ab}(l^{a}n^{b}+m^{a}\overline{m}^{b}), \\
\phi _{2} &=&F_{ab}n^{a}\overline{m}^{b}.
\end{eqnarray*}

From the radial asymptotic Bianchi identities and Maxwell equations, we have
their \textit{asymptotic behavior} (the 'peeling' theorem)\cite{NP}:%
\begin{eqnarray*}
\Psi _{0} &=&\Psi _{0}^{0}r^{-5}+O(r^{-6}), \\
\Psi _{1} &=&\Psi _{1}^{0}r^{-4}+O(r^{-5}), \\
\Psi _{2} &=&\Psi _{2}^{0}r^{-3}+O(r^{-4}), \\
\Psi _{3} &=&\Psi _{3}^{0}r^{-2}+O(r^{-3}), \\
\Psi _{4} &=&\Psi _{4}^{0}r^{-1}+O(r^{-2}).
\end{eqnarray*}%
\begin{eqnarray*}
\phi _{0} &=&\phi _{0}^{0}r^{-3}+O(r^{-4}), \\
\phi _{1} &=&\phi _{1}^{0}r^{-2}+O(r^{-3}), \\
\phi _{2} &=&\phi _{2}^{0}r^{-1}+O(r^{-2}),
\end{eqnarray*}%
with 
\begin{eqnarray*}
\Psi _{n}^{0} &=&\Psi _{n}^{0}(u,\zeta ,\overline{\zeta }), \\
\phi _{n}^{0} &=&\phi _{n}^{0}(u,\zeta ,\overline{\zeta }).
\end{eqnarray*}%
\qquad

The non-radial Bianchi Identities and Maxwell equations yield the evolution
equations for these leading terms (our basic variables): \ 
\begin{eqnarray}
\dot{\Psi}_{2}^{0\,} &=&-\eth \Psi _{3}^{0\,}+\sigma ^{0}\Psi
_{4}^{0\,}+k\phi _{2}^{0}\overline{\phi }_{2}^{0},  \label{AsyBI1} \\
\dot{\Psi}_{1}^{0\,} &=&-\eth \Psi _{2}^{0\,}+2\sigma ^{0}\Psi
_{3}^{0\,}+2k\phi _{1}^{0}\overline{\phi }_{2}^{0},  \label{AsyBI2} \\
\dot{\Psi}_{0}^{0\,} &=&-\eth \Psi _{1}^{0\,}+3\sigma ^{0}\Psi
_{2}^{0\,}+3k\phi _{0}^{0}\overline{\phi }_{2}^{0},  \label{AsyBI3} \\
k &=&2Gc^{-4},  \label{k}
\end{eqnarray}

\begin{eqnarray}
\dot{\phi}_{1}^{0\,} &=&-\eth \phi _{2}^{0},  \label{MaxI} \\
\dot{\phi}_{0}^{0\,} &=&-\eth \phi _{1}^{0}+\sigma ^{0}\phi _{2}^{0}.
\label{MaxII}
\end{eqnarray}

The u-derivative is denoted by the overdot.

After the final coordinate transformation to the static pseudo-Minkowski
coordinates and static pseudo-Minkowski cuts, the Eqs. (\ref{AsyBI1}-\ref%
{MaxII}) are seen to contain our classical (mechanical) equations of motion.

The quantity $\sigma ^{0}(u,\zeta ,\overline{\zeta })\ $(referred earlier to
as the asymptotic shear), is the leading term in the shear of the geodesic
congruence, $l^{a};$\ i.e.$.,$%
\begin{equation*}
\sigma =r^{-2}\sigma ^{0}(u,\zeta ,\overline{\zeta })+O(r^{-4}),
\end{equation*}%
while its first $u$-derivative$\ $is the Bondi news function. \ We consider $%
\sigma ^{0}(u,\zeta ,\overline{\zeta })\ $as a free function but take it
only up to the quadrupole terms, Eq.(\ref{asym shear}). It, as such, plays a
significant role in what later follows. From the spin-coefficient equations
one finds that

Body Math
\begin{eqnarray}
\Psi _{3}^{0} &=&\eth (\overline{\sigma }^{0})^{\cdot },  \label{sigma dot}
\\
\Psi _{4}^{0} &=&-(\overline{\sigma }^{0})^{\cdot \cdot }\ .  \notag
\end{eqnarray}

\subsection{Physical Identifications}

From the definition of the \textit{mass aspect,} ${\large \Psi ,}$ (real
from field equations) by%
\begin{equation}
\Psi =\overline{\Psi }\equiv \Psi _{2}^{0\,}+\eth ^{2}\overline{\sigma }%
^{0}+\sigma ^{0}(\overline{\sigma }^{0})^{\cdot },  \label{MassAspect}
\end{equation}

\noindent Bondi defined the asymptotic mass, $M_{B},$ and 3-momentum, $%
P_{B}^{i}\ \ $as the$\ l=0\ $\&$\ l=1\ $harmonic coefficients of $\Psi .\ $%
Specifically,

\textbf{Definition 1 \ }%
\begin{eqnarray}
\Psi &=&\Psi ^{0}+\Psi ^{i}Y_{1i}^{0}+\Psi ^{ij}Y_{2ij}^{0}+.  \label{DEF.1}
\\
\Psi ^{0} &=&-\frac{2\sqrt{2}G}{c^{2}}M_{B}  \label{mass} \\
\Psi ^{i} &=&-\frac{6G}{c^{3}}P_{B}^{i}  \label{momentum}
\end{eqnarray}

By rewriting Eq.(\ref{AsyBI1}), replacing the $\Psi _{2}^{0\,}$ by $\Psi $
via Eq.(\ref{MassAspect}), we have

\begin{equation*}
\dot{\Psi}\text{ }=\text{ }(\sigma ^{0})^{\cdot }(\overline{\sigma }%
^{0})^{\cdot }+\ k\phi _{2}^{0}\overline{\phi }_{2}^{0}.
\end{equation*}%
Immediately we have the Bondi mass/energy loss theorem: 
\begin{equation}
\dot{M}_{B}=-\frac{c^{2}}{2\sqrt{2}G}\int ((\sigma ^{0})^{\cdot }(\overline{%
\sigma }^{0})^{\cdot }+k\phi _{2}^{0}\overline{\phi }_{2}^{0})d^{2}S<0,
\label{BondiTheorem}
\end{equation}%
the integral taken over the unit 2-sphere at constant $\ u$. \ This
relationship is at the basis of almost all the contemporary work on the
detection of gravitational radiation.

\textbf{Definition\ \ 2 \ \ }Though it has been a controversial subject and
there is no general agreement, we \underline{\textit{adopt the definition}}
(which comes from linear theory) of the \textit{complex} mass dipole moment, 
$(D_{(complex)}^{i}=D_{(mass)}^{i}+ic^{-1}J^{i})$, as the $l=1\ $%
harmonic component of $\Psi _{1}^{0},\ $\ 
\begin{equation}
\Psi _{1}^{0}=-6\sqrt{2}Gc^{-2}(D_{(mass)}^{i}+ic^{-1}J^{i})Y_{1i}^{1}+....
\label{DEF.2}
\end{equation}%
$D^{i}\ $the mass dipole and $J^{i},\ $the total angular momentum, as "seen"
at null infinity. \ The main defense of this definition is that it works
extremely well.

\textbf{Definition\ \ 3 \ }Our identification - which is standard - for the
complex E\&M dipole, (electric and magnetic dipoles, $%
(D_{Elec}^{i}+iD_{Mag}) $) as the $l=1$ harmonic component of $\phi
_{0}^{0}\ $is: 
\begin{eqnarray}
\phi _{0}^{0} &=&2(D_{Elec}^{i}+iD_{Mag})Y_{1i}^{1},  \label{DEF.3} \\
D_{E\&M}^{i} &=&(D_{Elec}^{i}+iD_{Mag})=q\xi ^{i}.
\end{eqnarray}%
\qquad

Later we will connect these three physical identifications with the \textit{%
complex center of mass}.\textit{\ \ }

\textit{Actually, }for the general situation there is the independent
complex center of charge.\underline{ Here, however for simplicity, we 
\textit{assume that they coincide}}\emph{. \ This is not necessary but is a
simplifying restriction.\cite{ANK}}

\textbf{Comment. }For later use we note that from the asymptotic Maxwell
equations, Eqs.(\ref{MaxI}) and (\ref{MaxII}), we have that\cite{ANK}

\begin{eqnarray}
\phi _{1}^{0} &=&q+\sqrt{2}q\xi ^{i\ \prime }Y_{1i}^{0}+Q_{1}+...
\label{MaxIII} \\
\phi _{2}^{0} &=&-2q\xi ^{i\ \prime \prime }Y_{1i}^{-1}+Q_{2}+...,  \notag
\end{eqnarray}%
with the $Q\ $s representing known quadrupole terms.

\textbf{Comment. }The indices 0,i,j,k.... have direct geometric meaning
coming from the position or tangent vectors in $H$-space. They also can be
interpreted as vectors in a representation space of the Lorentz group that
has its origin via a subgroup of the BMS group acting on $\mathcal{I}^{+}$.%
\cite{ANK}

\subsection{Modus Operandi}

Our operation now consists of taking the Weyl tensor components (mainly $%
\Psi _{1}^{0\,}$and $\Psi _{2}^{0\,}$) and transferring them from the Bondi
tetrad ($l^{a},\overline{m}^{a},m^{a},n^{a}$) to the pseudo-Minkowski tetrad
($l^{\ast a},m^{\ast a},\overline{m}^{\ast a},n^{\ast a}$) (actually doing
this two times) via%
\begin{eqnarray}
l^{\ast a} &=&l^{a}+b\overline{m}^{a}+\overline{b}m^{a}+0(r^{-2}),
\label{null rot II} \\
m^{\ast a} &=&m^{a}+bn^{a},  \notag \\
n^{\ast a} &=&n^{a},  \notag \\
b &=&-\frac{L}{r}+0(r^{-2}).  \notag
\end{eqnarray}%
and

\begin{eqnarray}
\Psi _{0}^{\ast 0} &=&\Psi _{0}^{0}-4L\Psi _{1}^{0}+6L^{2}\Psi
_{2}^{0}-4L^{3}\Psi _{3}^{0}+L^{4}\Psi _{4}^{0},  \label{0} \\
\Psi _{1}^{\ast 0} &=&\Psi _{1}^{0}-3L\Psi _{2}^{0}+3L^{2}\Psi
_{3}^{0}-L^{3}\Psi _{4}^{0},  \label{1} \\
\Psi _{2}^{\ast 0} &=&\Psi _{2}^{0}-2L\Psi _{3}^{0}+L^{2}\Psi _{4}^{0},
\label{2} \\
\Psi _{3}^{\ast 0} &=&\Psi _{3}^{0}-L\Psi _{4}^{0},  \label{3} \\
\Psi _{4}^{\ast 0} &=&\Psi _{4}^{0}.  \label{4}
\end{eqnarray}

The $L\ $and its complex conjugate, $\overline{L},$ are determined by Eq.(%
\ref{L&Lbar}) with $G\ $given\ by, Eq.(\ref{cuts*})

\begin{equation}
G=\frac{\tau }{\sqrt{2}}-\frac{1}{2}\xi ^{i}(\tau )Y_{1i}^{0}(\zeta ,%
\overline{\zeta })+\xi ^{ij}(\xi ^{a}(\tau ))Y_{ij}^{2}(\zeta ,\overline{%
\zeta })+..  \label{cuts II}
\end{equation}%
We perform this tetrad rotation twice: first with the dynamic
pseudo-Minkowski coordinates, the $\xi ^{a}(\tau )\ $to be determined by a
center of mass condition and then again for the static pseudo-Minkowski
coordinates, $\xi ^{a}=\tau \delta _{0}^{a}\ $done to put the final results
in a pseudo-Lorentzian frame.

In addition, in each case, we must change the $I^{+}\ $coordinates from
Bondi to the ($\tau ,\zeta ,\overline{\zeta }$)\ slicing via Eq.(\ref{cuts}).

\textit{Aside: Eventually the} ($\tau ,\zeta ,\overline{\zeta }$)\textit{\
will be changed to the real }$I^{+}\ $\textit{coordinates (}$t,\zeta ,%
\overline{\zeta }$\textit{).}

These transformations will be implemented in several stages.

\textbf{Stage I}. We go from the Bondi coordinates to the pseudo-Minkowski
situation via Eq.(\ref{cuts*}). After both the tetrad and coordinate change,
for constant $\tau ,\ $we concentrate on the $l=1\ \ $component, i.e., $\Psi
_{1}^{\ast 0i}\ $of $\Psi _{1}^{\ast 0}.\ $Using$\ $the definition of the 
\textit{complex} mass dipole moment, the complex center of mass `position'
is determined (defined) by setting the $\Psi _{1}^{\ast 0i}$ equal to zero.
This allows us to determine the three components of $\xi ^{i},\ $(the center
of mass)$,\ $with$\ \xi ^{0}=\tau .$

\textbf{Stage II. }With the\textbf{\ }$\xi ^{a}(\tau )\ $now known we can go
back and find the $l=1\ $spherical\ harmonic$\ $component of the Bondi $\Psi
_{1}^{0i}(u)$ in terms of the complex center of mass. Using, in the Bondi
frame, the evolution equations, Eqs.(\ref{AsyBI1})-(\ref{MaxII}), with the
just found center of mass position, $\xi ^{a}(u),\ $many of our mechanical
equations are obtained but now expressed as functions of our $u$.

\textbf{Stage III. }With all these relations, including the complex center
of mass and the dynamics - expressed in the Bondi frame - we do the
transformation back to the pseudo-Minkowski system but now to the \emph{%
static frame} to obtain our final results\textbf{. }The "\textit{static frame%
}" mimics the ordinary flat-space Lorentzian frame.

The calculations involved in these three stages were rather involved. They
included the coordinate and tetrad transformations between the Bondi frame
and the pseudo-Minkowski frames several times, they often involved Taylor
expansions up to the quadrupole terms and the frequent use of Clebsch-Gordon
expansion of the spherical harmonics products. \ Since much of this has been
completed and appeared in published and refereed literature,\cite{ANK}, we
will not redo them but simply refer to these results for our present use.

\section{\ Results}

\subsection{\textbf{Stage I }}

We begin with an asymptotically flat space-time in a Bondi coordinate and
tetrad system (previous section) and perform the null rotation, Eq.(\ref%
{null rot II}), to the shear-free null vector field%
\begin{eqnarray}
l^{\ast a} &=&l^{a}+b\overline{m}^{a}+\overline{b}m^{a}+0(r^{-2}),
\label{rot 2} \\
m^{\ast a} &=&m^{a}+bn^{a},  \notag \\
n^{\ast a} &=&n^{a},  \notag \\
b &=&-\frac{L}{r}+0(r^{-2}).  \notag
\end{eqnarray}%
with $L=\eth G(\xi ^{a}(\tau ),\zeta ,\overline{\zeta }),$ and its complex
conjugate,$\ \overline{L}.\ $The $G\ \ $is given by Eq.(\ref{cuts II}),

\begin{equation}
u=G=\frac{\tau }{\sqrt{2}}-\frac{1}{2}\xi ^{i}(\tau )Y_{1i}^{0}(\zeta ,%
\overline{\zeta })+\xi ^{ij}(\xi ^{a}(\tau ))Y_{ij}^{2}(\zeta ,\overline{%
\zeta })+..  \label{cuts II*}
\end{equation}%
with $\xi ^{a}(\tau )\ \ $an unknown world-line in $H$-space - to be
determined. The relevant (for us) Weyl tensor component, $\Psi _{1}^{0},\ $%
transforms, Eq.(\ref{1}), as

\begin{equation}
\Psi _{1}^{\ast 0}=\Psi _{1}^{0}-3L\Psi _{2}^{0}+3L^{2}\Psi
_{3}^{0}-L^{3}\Psi _{4}^{0}.  \label{PSI10}
\end{equation}%
Considering $L$ and $\sigma ^{0}\ $as first order and $M_{B}\ $in Eq.(\ref%
{MassAspect}), as zero order, with the Mass Aspect $\Psi ,$

\begin{eqnarray}
\Psi &=&\Psi ^{0}+\Psi ^{i}Y_{1i}^{0}+\Psi ^{ij}Y_{2ij}^{0}+.  \label{M.A.1}
\\
\Psi ^{0} &=&-\frac{2\sqrt{2}G}{c^{2}}M_{B}  \notag \\
\Psi ^{i} &=&-\frac{6G}{c^{3}}P_{B}^{i}  \notag
\end{eqnarray}%
$\ $ we have

\begin{equation}
\Psi _{1}^{\ast 0}=\Psi _{1}^{0}-3L(\Psi -\eth ^{2}\overline{\sigma }^{0}).
\label{PSI10*}
\end{equation}

Our procedure for finding the complex center of mass now centers on Eq.(\ref%
{PSI10*}). The right-side, which is a function of both $u\ $and $\tau ,$ is
transformed to a function of only $\tau \ $via Eq.(\ref{cuts II*}).$\ \ $All
the variables on the right side are then expanded in spherical harmonics and
simplified by Clebsch-Gordon expansions. We separate out and \underline{%
\textit{set to zero}} the $l=1\ $harmonics on the right side,\ i.e., we
force $\Psi _{1}^{\ast 0i}$ $=0\ $for constant $\tau $ slices. This is a
lengthy and difficult task and approximations are needed.

From this result there are two things that could be done: (1) from the three 
$l=1$ terms we could solve for the three $\xi ^{i}(\tau )\ $or, as we do,
(2) go backwards and solve for the original $\Psi _{1}^{0i}\ $but now in
terms of the $\xi ^{i}(\tau ).\ $

After considerable work we have $\Psi _{1}^{0}\ ,\ $found$\ $first as a
function of $\tau \ $and then finally, via the (approximate) inverse
function to Eq.(\ref{cuts II*}), i.e.,%
\begin{equation*}
\frac{\tau }{\sqrt{2}}=u+\frac{1}{2}\xi ^{i}(u)Y_{1i}^{0}(\zeta ,\overline{%
\zeta })+\xi ^{ij}(\xi ^{a}(u))Y_{ij}^{2}(\zeta ,\overline{\zeta })+..,
\end{equation*}
then expressed as a function of $u.\ $After its spherical harmonic
expansion, we finally have the $l=1$ harmonic coefficient,$\ \Psi _{1}^{0i}$
\ -\ (needed for our definition 2, Eq.(\ref{DEF.2}), of the complex mass
dipole), namely

\begin{eqnarray*}
\Psi _{1}^{0i} &=&-\frac{6\sqrt{2}G}{c^{2}}M_{B}\xi ^{i}+i\frac{6\sqrt{2}G}{%
c^{3}}P_{B}^{k}\xi ^{j}\epsilon _{kji}-\frac{576G}{5c^{3}}P_{B}^{k}\xi
^{ik}+i\frac{6912\sqrt{2}}{5}\xi ^{lj}\overline{\xi }^{lk}\epsilon _{jki} \\
&&-i\frac{2\sqrt{2}G}{c^{6}}q^{2}\xi ^{k}\overline{\xi }^{j\prime \prime
}\epsilon _{kji}-\frac{48G}{5c^{6}}q^{2}\xi ^{ji}\overline{\xi }^{j\prime
\prime }-\frac{4G}{5c^{7}}q^{2}\xi ^{j}\overline{Q}_{C}^{ij\prime \prime
\prime }-i\frac{16\sqrt{2}G}{5c^{7}}q\xi ^{lj}\overline{Q}_{C}^{lk\prime
\prime \prime }\epsilon _{jki}.
\end{eqnarray*}

Considering that many of the quadrupole terms involve high powers of$\
c^{-1}\ $and we can consider quadrupole-quadrupole interactions as weak, we
approximate $\Psi _{1}^{0i}\ $simply$\ $as

\begin{equation}
\Psi _{1}^{0i}(u)=-\frac{6\sqrt{2}G}{c^{2}}M_{B}\xi ^{i}+i\frac{6\sqrt{2}G}{%
c^{3}}P_{B}^{k}\xi ^{j}\epsilon _{kji}\ ...  \label{***}
\end{equation}

\subsection{Stage 2}

Equating our \textit{definition} 2 
\begin{equation}
\Psi _{1}^{0i}=-6\sqrt{2}Gc^{-2}(D_{(mass)}^{i}+ic^{-1}J^{i})...
\label{PSI10i&}
\end{equation}%
of the complex mass dipole with Eq.(\ref{PSI10i&}), using 
\begin{equation}
\xi ^{i}=\xi _{R}^{i}+i\xi _{I}^{i},  \label{real&complex}
\end{equation}%
we obtain our

\textbf{Result:1}

\begin{eqnarray}
D_{(mass)}^{i} &=&M_{B}\xi _{R}^{i}-c^{-1}P_{B}^{k}\xi _{I}^{j}\ \epsilon
_{jki}+...,  \label{mass dipole} \\
J^{i} &=&cM_{B}\xi _{I}^{i}+P_{B}^{k}\xi _{R}^{j}\epsilon _{jki}+....
\label{ang mom}
\end{eqnarray}%
or 
\begin{eqnarray}
\overrightarrow{D}_{(mass)} &=&M_{B}\overrightarrow{r}+c^{-2}M_{B}^{-1}%
\overrightarrow{P}_{B\ }\mathrm{x}\overrightarrow{S}.  \label{D} \\
\overrightarrow{r} &=&\xi _{R}^{i}=(\xi _{R}^{1},\xi _{R}^{2},\xi _{R}^{3}),
\\
\overrightarrow{S} &=&cM_{B}\xi _{I}^{j}=cM_{B}(\xi _{I}^{1},\xi
_{I}^{2},\xi _{I}^{3}), \\
\overrightarrow{J} &=&\overrightarrow{S}+\overrightarrow{r}\mathrm{x}%
\overrightarrow{P}.
\end{eqnarray}%
\ 

The first term in $D_{(mass)}^{i}\ $is the standard dipole definition while
the second term is identical to a dipole term in the relativistic
angular-momentum tensor\cite{Enigma}\cite{AngMom}.

The expression for angular momentum has (1), the intrinsic spin $%
\overrightarrow{S}$ (same as for the Kerr metric)\cite{ANK} and (2) the
conventional orbital angular momentum term.

We find the details, arising from the definitions 1 and 2 and the Einstein
Equations, to be rather remarkable - and strange. \ Note that the $z^{a}\ $%
are $H$-space coordinates.\ 

Continuing, we substitute Eq.(\ref{PSI10i&}) into the evolutionary Bianchi
Identity, Eq.(\ref{AsyBI2}),

\begin{equation}
\dot{\Psi}_{1}^{0\,}=-\Psi _{2}^{0\,}+2\sigma ^{0}\Psi
_{3}^{0\,}+2k\phi _{1}^{0}\overline{\phi }_{2}^{0},  \label{BIII}
\end{equation}%
and using \textbf{definition} 1 and Eq.(\ref{MaxIII}), we find, directly
from the real part (with no manipulation), that

\textbf{Result: 2}%
\begin{eqnarray}
P_{B}^{i} &=&M_{B}\ \xi _{R}^{i\ \prime }-\frac{2q^{2}}{3c^{3}}\xi _{R}^{i\
\prime \prime }+H.O.  \label{P=Mv} \\
H.O. &=&\text{quadrupole and higher order terms.}
\end{eqnarray}

The prime denotes the u-derivative.

We obtain, for the Bondi momentum, the familiar kinematic $M_{B}%
\overrightarrow{v}$ term and a term familiar from electrodynamics, the
radiation reaction contribution to the linear momentum.

From the imaginary part of the Bianchi Identity we have the momentum loss
equation;

\textbf{Result: 3}%
\begin{equation}
J^{i\ \prime }=-\frac{2q^{2}}{3c^{3}}\xi _{I}^{i\ \prime \prime }+\frac{%
2q^{2}}{3c^{3}}(\xi _{R}^{j\ \prime }\xi _{R}^{k\ \prime \prime }+\xi
_{I}^{k\ \prime }\xi _{I}^{k\ \prime \prime })\epsilon _{kji}+H.O.\text{ }
\label{J prime}
\end{equation}

There are several things to note: (1) the first term on the right side can
be moved to the left, simply changing the definition of $J^{i}\ \ $to 
\begin{equation*}
J^{\ast i}=J^{i}+\frac{2q^{2}}{3c^{3}}\xi _{I}^{i\ \prime },
\end{equation*}%
i.e., adding an electromagnetic part to the spin term $S^{i},\ $and$\ $(2)
Landau \& Lifschitz \cite{LL} have a special case of Eq.(\ref{J prime}).
They omit the $\xi _{I}^{i}\ $terms.

Finally substituting the relevant terms into the evolutionary Bianchi
Identity, Eq.(\ref{AsyBI1}), 
\begin{subequations}
\begin{equation}
\dot{\Psi}_{2}^{0\,}=-\Psi _{3}^{0\,}+\sigma ^{0}\Psi
_{4}^{0\,}+k\phi _{2}^{0}\overline{\phi }_{2}^{0},  \label{BI1*}
\end{equation}%
we have, first for the $l=0$\ harmonic coefficient, the (Bondi) mass loss
expression but now including the well known (classical) electromagnetic
energy losses, i.e.,

\textbf{Result: 4} 
\end{subequations}
\begin{eqnarray}
M_{B}^{\prime } &=&-\frac{G}{5c^{7}}(Q_{Mass}^{jk\prime \prime \prime
}Q_{Mass}^{jk\prime \prime \prime }+Q_{Spin}^{jk\prime \prime \prime
}Q_{Spin}^{jk\prime \prime \prime })-\frac{4q^{2}}{3c^{5}}(\xi _{R}^{i\prime
\prime }\xi _{R}^{i\prime \prime }+\xi _{I}^{i\prime \prime }\xi
_{I}^{i\prime \prime }).  \label{mass loss} \\
&&-\frac{4}{45c^{7}}(Q_{E}^{jk\prime \prime \prime }Q_{E}^{jk\prime \prime
\prime }+Q_{M}^{jk\prime \prime \prime }Q_{M}^{jk\prime \prime \prime })
\end{eqnarray}%
\noindent \noindent \noindent The first term is the standard Bondi
quadrupole mass loss (now including the \textit{spin quadrupole}
contribution to the loss - maybe new), the second and third terms are the
classical E\&M dipole and quadrupole energy loss - including the correct
numerical factors.

The $l=1\ $terms lead to the momentum loss expression,

\textbf{Result: 5}

\begin{equation}
P_{B}^{i\ \prime }=F_{recoil}^{i}  \label{P'}
\end{equation}%
where $F_{recoil}^{i}\ $is composed of many non-linear radiation terms
involving the time derivatives of the gravitational quadrupole and the E\&M
dipole and quadrupole moments. These terms are known and given\cite{ANK} but
not relevant to us now. \ Instead we substitute Eq.(\ref{P=Mv}) into Eq.(\ref%
{P'}) leading to Newton's second law;

\begin{equation}
M_{B}\xi _{R}^{i\prime \prime }=F^{i}\equiv M_{B}^{\prime }\xi _{R}^{i\prime
}+\frac{2q^{2}}{3c^{3}}\xi _{R}^{i\prime \prime \prime }+F_{recoil}^{i}.
\label{F=ma}
\end{equation}%
\textbf{Result: 6}\qquad \qquad \qquad

We find this rather astonishing - there is exactly the classical mechanics
standard rocket mass loss expression and the classical radiation reaction
term. \ The last term is just the momentum recoil force - also known
explicitly.

In the context of these results we mention two further automatic results:

1. From earlier results,$\ \ $%
\begin{eqnarray*}
\ \xi _{R}^{i} &=&center\ of\ mass\ position \\
S^{i} &=&Mc\xi _{I}^{i}\ =\ Spin-Angular\ momentum \\
\ D_{M}^{i} &=&q\xi _{I}^{i}=\ Magnetic\ dipole\ Moment
\end{eqnarray*}%
and the classical classical gyromagnetic ratio $\gamma =\frac{q}{2Mc}g,\ $we
see 
\begin{equation}
\gamma =\frac{D_{M}^{i}}{L_{spinang.mom}}=\frac{q\xi _{I}^{i}}{M_{B}c\xi
_{I}^{i}}=\frac{q}{M_{B}c},  \label{g factor}
\end{equation}%
and discover the Dirac value of the $g$-factor to be $g$=$2$.

2. In classical relativistic mechanics \cite{AngMom} one has the definition
of the relativistic angular momentum tensor, $M^{ab}$,%
\begin{eqnarray*}
M^{ab} &=&L^{ab}+S^{ab} \\
L^{ab} &=&2MX^{[a}V^{b]} \\
S^{ab} &=&-\eta ^{abcd}S_{c}^{\ast }V_{d},\ \ \ S_{c}^{\ast }V^{c}=0
\end{eqnarray*}%
so that%
\begin{eqnarray}
M^{ij} &=&L^{ij}+S^{ij}  \label{ang mom 2} \\
&=&M(X^{i}V^{j}-V^{i}X^{j})-\epsilon ^{ijk}(S_{k}^{\ast
}V_{0}-V_{k}S_{0}^{\ast })  \notag
\end{eqnarray}

Using our results, $S^{i}=cS^{\ast i}=Mc\xi _{I}^{i},\ S_{0}=0,\ V_{0}\sim
1,\ V_{k}\sim 0\ $and multiplying by $\epsilon _{ijk},\ $we have agreement
with our Eq.(\ref{ang mom}).

Then from%
\begin{eqnarray}
M^{0i} &=&L^{0i}+S^{0i}  \label{mass dipole 2} \\
&=&2MX^{[0}V^{i]}-\eta ^{0ijk}S_{j}^{\ast }V_{k}  \notag \\
&=&M_{B}\xi _{R}^{i}-\epsilon _{ijk}c^{-1}\xi _{I}^{j}P^{k},  \notag
\end{eqnarray}

\noindent we have agreement with our Eq.(\ref{mass dipole}), i.e., with our $%
\overrightarrow{P}\mathrm{x}\overrightarrow{S}\ $term.

The \textit{relativistic angular momentum tensor} (unrelated to physical
space-time) is\underline{ \textit{sitting quietly and unobserved in our Weyl
tensor}}.

\subsection{Stage 3}

Going from the Bondi slicings of stage 2 to the "static" real
pseudo-Minkowski slicing of stage 3 is easy. From Eq.(\ref{case 2}), we see
that the $l=1\ $\ harmonics are now missing

\begin{eqnarray}
\widehat{\tau } &=&\widehat{t}+i\widehat{\lambda }  \label{case 2*} \\
\widehat{\lambda } &=&\widehat{\Lambda }(\widehat{t},\zeta ,\overline{\zeta }%
)\equiv -\sqrt{2}\xi _{I}^{ij}(\widehat{t})Y_{2ij}^{0}(\zeta ,\overline{%
\zeta }),  \notag \\
u_{R} &=&\frac{\widehat{t}}{\sqrt{2}}+\xi _{R}^{ij}(\widehat{t}%
)Y_{2ij}^{0}(\zeta ,\overline{\zeta })...  \notag
\end{eqnarray}

From our approximations, i.e., disregarding the quadrupoles, we can simply
replace our $u\ $\ with $\widehat{t},\ $so that all our \textbf{Results, }%
1-6, remain unchanged. We emphasize that these \textbf{Results} are for real
time and real coordinates even though most of the calculations involved
complex variables.

\section{VI. Discussion}

Several Remarks are in order.

1. The present work is the third in a developing series of papers dealing
with unusual structures hiding on $\mathcal{I}^{+}.\ $In the earlier work%
\cite{ANK}\cite{Enigma} the coordinates as well as the time (for the
classical equations of mechanics), were, in general complex - here they are
real. \ Furthermore the $\mathcal{I}^{+}\ $cuts were complex and were open
to the perplexing question "What was the significance of the complex Bondi
slicing? Why was their use important?" Here we have the answer. They were
not important. The important observation is that here we are using slicings
(pseudo-Minkowskian) that mimic (or are as close to) real flat-Lorentzian
slicing as possible - but give the same results as in the Bondi slicing.
This does clear away one of the enigmas associated with the earlier work\cite%
{Enigma} - why was Bondi slicing important?

2. It is the collection of results, Eqs. (\ref{mass dipole}), (\ref{ang mom}%
), (\ref{P=Mv}),(\ref{J prime}),(\ref{mass loss}) and (\ref{F=ma}),
mimicking or imitating classical mechanics, that constitute our main
contribution - in conjunction with our novel pseudo-Minkowskian slicing.
Nevertheless there still remains the major enigma, "Why and how is it that
the (now) real $H$-space coordinates play so accurately the role of
space-time coordinates. Though they appear to be space-time equations of
motion there is \emph{no physical space-time that is associated with the
equations}. They (as we said earlier) take place on the strange $H$-space. \
Is this just a giant coincidence? We find that difficult to believe. \ What
possible meaning can one give to them - and to the $H$-space? \ What is the
\ physical meaning of the decomposition, $\xi ^{i}(\tau )=\xi _{R}^{i}(\tau
)+i\xi _{I}^{i}(\tau ).\ $Why should the imaginary part of the $H$-space
coordinate be related to spin? Is there something here of real significance?
We do not know - but it is suggestive.

We also observe the curious result - in our construction - that when we have
spin-angular momentum, we also have the case of the twisting sear-free
congruence. Twist seems happily to be related to spin.

3. There is the suggestion that $H$-space might play the role of some sort
of observation or optical image space. Due to curvature affects one can not
look straight back and expect to see distant objects along the line of
sight. Is this a manifestation of this image forming? This is pure
speculation and not at all clear.

4. Among the 6 \textbf{Results} of the previous section, all but\ \textbf{%
Result} \#3 are identical to standard classical (mechanical and E\&M)
equations. The special case of \textbf{Result} \#3 where spin is omitted is
given in reference (\cite{LL}). We have\ not been able to find any reference
that contains our complete result. \ A related question is - if our
conservation of angular momentum result is new, can we consider it to be a
prediction.

5. We point out the utter simplicity of the origin of the radiation reaction
term in the expression for $P_{B}^{i}$.\ \ In Eq.(\ref{AsyBI2})%
\begin{equation*}
\dot{\Psi}_{1}^{0\,}=-\eth \Psi _{2}^{0\,}+2\sigma ^{0}\Psi
_{3}^{0\,}+2k\phi _{1}^{0}\overline{\phi }_{2}^{0},
\end{equation*}%
the$\ P_{B}^{i}$ $,$\ is sitting in the $\eth \Psi _{2}^{0\,},\ $the $q\ $%
sits in the $\phi _{1}^{0}$\ , while $\xi ^{i\ \prime \prime }\ \ $is in the 
$\overline{\phi }_{2}^{0}.$ \ The numerical factors are there. The radiation
reaction term is just sitting there - no assumptions, no derivation, it is
just waiting to be observed in the equation.

6. \ We emphasize several things;

a. We are using only the standard Einstein equations coupled in the standard
way to the standard Maxwell equations. Furthermore we are using the standard
asymptotically flat solutions of Bondi,\ Sachs, Penrose and Newman-Unti -
with nothing further added. \ We note that in this standard framework we
search for shear-free null geodesic congruences via the so-called "Good Cut
Equation" - our main research tool. \ From the null surfaces associated with
these congruences and their intersections with the asymptotically flat $%
\mathcal{I}^{+}$, we construct -(mimicking the virtually identical
flat-space construction) the one-parameter family of real cuts.

b.\ On these cuts, using the tetrad adapted to the cuts (different from the
Bondi tetrad), we \underline{\textit{defined}} the complex center of mass in
terms of a Weyl tensor component. \ From that definition and that of the
Bondi mass and momentum, all of our \textbf{Results} followed. \ No formal
lengthy derivations were needed; the \textbf{Results} were just sitting
there in the Weyl tensor and Bianchi Identities.

c. The subtlety of some of the results were rather surprising: e.g., finding
the rocket force, the Abraham-Lorentz radiation reaction force as well as
the electromagnetic energy loss from both the dipole and quadrupole
radiation (both of electric and magnetic type, and all with correct
coefficients) and the angular momentum loss. \ Some of the results clearly
arise from the inclusion of the Maxwell stress-energy tensor in the
Einstein-Maxwell equations. There however was no straight matter stress
tensor\ used.

7. Though it is not obvious where to go next, we notice that there is a
chance to expand on Penrose's Asymptotic Twistor theory via twistor curves
on the new family of shear-free cuts of \ $\mathcal{I}^{+}.\ $In the past
they were defined on the associated $H$-space. \ Now it could possibly be
done on the physical space itself. This must be investigated.

8. The natural question also arises - can our method of obtaining equations
of motion be extended and applied to a two-body problem? At the moment we
see possibilities but not good ones.

9. \ We have not yet explored how the BMS group interacts with the present
results - via representation theory.

10. And finally we fully acknowledge that we have no good idea for the
physical meaning of the $H$-space and its coordinates -- why their real
parts should be appropriate for the description of the center-of-mass motion
or why their imaginary parts are associated with spin? Does the $H$-space
metric play any role? Is all this an empty accident? or is there something
profound? It is sufficiently crazy and far-off the mainstream that it might
well be profound. \ In this context, we note that instead of looking at the
leading Weyl tensor terms, we could - getting the same results - just as
well used weighted integrals (powers of $r\ $and spherical harmonics) over
the\ "shear-free" spheres (cuts) at $\mathcal{I}^{+}\ $that so resemble
Minkowski space light-cone cuts. \ This seems to help pick out our
Lorentzian-like mechnical results.

\section{VII. Acknowledgements}

We gratefully thank Timothy Adamo for both a careful critical reading of the
manuscript, for hours of enlightening discussions and collaboration on an
earlier manuscript\cite{ANK} where many of the present ideas were developed.
We are deeply indebted to Roger Penrose for years of friendship and patient
explanations and encouragement on exploring the present ideas.

\bigskip

\bigskip

\bigskip

\bigskip

\end{document}